\documentclass[12pt,preprint]{aastex}

\usepackage{emulateapj5}

\begin{document}
\title{The HD 163296 Circumstellar Disk in Scattered Light: Evidence of 
Time-Variable Self-Shadowing}
\author{John P. Wisniewski\altaffilmark{1,2}, Mark Clampin\altaffilmark{1}, 
Carol A. Grady\altaffilmark{1,3}, David R. Ardila\altaffilmark{4}, Holland C. 
Ford\altaffilmark{5}, David A. Golimowski\altaffilmark{5}, Garth D. 
Illingworth\altaffilmark{6}, \& John E. Krist\altaffilmark{7}}

\altaffiltext{1}{NASA Goddard Space Flight Center, Code 667, 
Greenbelt, MD 20771 John.P.Wisniewski@gmail.com, Mark.Clampin@nasa.gov, 
Carol.A.Grady@nasa.gov}
\altaffiltext{2}{NPP Fellow}  
\altaffiltext{3}{Eureka Scientific, 2452 Delmer Street Ste 100, Oakland, CA 94602}
\altaffiltext{4}{Spitzer Science Center, Pasadena, CA 91125, ardila@ipac.caltech.edu}
\altaffiltext{5}{Johns Hopkins University, 3400 North Charles Street, Baltimore, MD 20218, 
ford@pha.jhu.edu,dag@pha.jhu.edu}
\altaffiltext{6}{Lick Observatory, University of California, Santa Cruz, CA 95064, gdi@ucolick.org}
\altaffiltext{7}{Jet Propulsion Laboratory, 4800 Oak Grove Drive, MS183-900, Pasadena, CA 
91109, john.e.krist@jpl.nasa.gov}

\begin{abstract} We present the first multi-color view of the scattered 
light disk of the Herbig Ae star HD 163296, based on coronagraphic observations from the
\textit{Hubble Space Telescope} Advanced Camera for Surveys (ACS).  
Radial profile fits of the surface brightness along the disk's semi-major axis indicates 
that the disk is not continuously flared, and extends to $\sim$540 AU.  The disk's 
color (V-I)=1.1 at a radial distance of 3$\farcs$5 is redder than the observed stellar color 
(V-I)=0.15.  This red 
disk color might be indicative of either an evolution in the grain size distribution (i.e. grain 
growth) and/or composition, both of which would be consistent with the observed 
non-flared geometry of the outer disk.  We also identify a single ansa 
morphological structure in our F435W ACS data, which is absent from 
earlier epoch F606W and F814W ACS data, but corresponds to one of the two ansa 
observed in archival HST STIS coronagraphic data.

Following transformation to similar band-passes, we find that the scattered light disk of HD 163296  
is 1 mag arcsec$^{-2}$ fainter at 3$\farcs$5 in the STIS data than in the ACS data.  Moreover, 
variations are seen in (i) the visibility of the ansa(e) structures, in (ii) the relative surface brightness 
of the ansa(e) structures, and in (iii) the (known) intrinsic polarization of the system.  These results 
indicate that the scattered light from the HD 163296 disk is variable.  We 
speculate that the inner disk wall, which Sitko et al. suggests has a variable scale height as diagnosed 
by near-IR SED variability, induces variable self-shadowing of the outer disk.  We further 
speculate that the observed surface brightness variability of the ansa(e) structures 
may indicate that the inner disk wall is azimuthally asymmetric. 
  
\end{abstract}


\section{Introduction} \label{intro}

Investigations of the nature and evolution of dust grains in proto-planetary systems are  
motivated in part by our desire to understand the birth and evolution of planetary bodies, 
which originate from these systems.  It is well established that Herbig 
Ae/Be stars \citep{her60} are 
intermediate mass pre-main-sequence stars, analogous to the more familiar low mass T Tauri 
stars, which contain copious amounts of circumstellar gas and dust
\citep{wat98}.  A wealth of observational evidence \citep{man97,oud99,man00,
vin02,eis04} 
suggests that the geometry of this gas and dust takes the form of a circumstellar disk.
By contrast, our understanding of fundamental properties such as grain composition and 
size distribution, as well as the time evolution of these parameters, are much less understood.

Analysis of ISO spectra led \citet{mee01} to develop a 2-part evolutionary classification 
of Herbig systems;  
Group I sources were characterized as being slightly younger systems with flared outer 
disks, while Group II sources were characterized as slightly more 
evolved systems which have flatter disks 
owing to grain growth and settling.  While \citet{mee01} invoked dust settling to explain 
the differences in flaring between Group I and II sources, it has also been suggested 
that the outer regions of some Herbig disks may experience self-shadowing, 
owing to a ``puffed-up'' inner circumstellar disk rim \citep{nat01,dul01,dul04,ise05}.  
Recent models which include inflated inner disk rims have proven to be successful in 
explaining the near-IR and interferometric properties of several Herbig Ae stars \citep{ise06}.  Conclusive confirmation of the geometry of the outer disk regions of 
Herbig Ae/Be stars inferred by SED-based studies has generally not occurred.

HD 163296 is a young (4 Myr; \citealt{van00}), nearby (122 pc; \citealt{van98}) 
Herbig Ae star.  While the star is not deeply embedded in a natal star formation 
dust cloud \citep{the85}, it still displays clear signs of active accretion via the presence 
of jets and Herbig Haro knots \citep{gra00,dev00,was06}.  Hence, HD 163296 appears 
to be in a transition phase between optically thick, extremely young pre-main-sequence 
stars and the much more optically thin, near zero-age-main-sequence debris disk type 
stars \citep{ard04,kal05,kal06,gol06}.  Indirect techniques \citep{bjo95} and resolved 
imaging \citep{man97,gra00,ise07} have confirmed the presence of a circumstellar disk associated 
with HD 163296.  HST STIS white-light coronagraphic observations spatially resolved the outer 
regions of this disk, and detected evidence of disk structure including an inner annulus of 
reduced scattering and a bright outer ring or ansae of material \citep{gra00}.  The 
inner region of HD 163296's 
circumstellar disk, as diagnosed by near-IR (0.8 - 5 $\mu$m) SED monitoring, exhibits 
evidence of variability, possibly owing to changes in the inner disk wall \citep{sit06}.

Attempts to link processes which dominate the inner regions of Herbig Ae disks to 
morphological features observed in the outer disk regions, i.e. determining the relative 
roles of self-shadowing, dust 
settling, and disk flaring \citep{dul04}, requires a wealth of observational data which diagnose
the inner and outer disk regions.  In this paper, we present a multi-epoch, multi-color 
view of the HD 163296 scattered light disk, providing a significant improvement in the 
documented behavior of the outer regions of HD 163296's circumstellar disk.  We describe the 
observational details and data reduction procedures applied to our ACS data in 
Section \ref{obs-red}.  In Section \ref{acs-charact}, we document the basic 
features of the HD 163296 scattered light disk in our ACS data, and compare these results 
to earlier epoch scattered light imaging in Section \ref{stis}.  The possible mechanisms 
which might explain the observed scattered light variability are discussed in 
Section \ref{origin}.  We provide a summary of our main results in Section \ref{summary}.

\section{Observations and Data Reduction} \label{obs-red}

We obtained coronagraphic imaging of HD 163296 in the HST ACS F435W filter via the
cycle 12 GTO program 9987, and in the ACS F606W and F814W filters via the cycle 11
GTO program 9295.  All observations utilized 
the coronagraphic mode of the High Resolution 
Channel (HRC) camera, along with its 1$\farcs$8 diameter occulting spot, yielding a 
29$\farcs$2 x 26$\farcs$2 field of view at a pixel scale of 0$\farcs$028 x 
0$\farcs$025.  The FWHM of point spread function (PSF) of the HRC is 50 mas in V.  
Our science target data were obtained at two different HST roll angles to  
better facilitate the separation of PSF artifacts from real structures.  The 
A3IV star HD 145570, selected for its spectral match to HD 163296, 
was used as a PSF reference, and was observed  
in orbits immediately following our HD 163296 exposures.  A summary of the 
basic parameters of our dataset is presented in Table \ref{datatable}.  

The standard HST pipeline was used to perform basic initial data reduction 
procedures.  We used the best available calibration files for this processing, 
including a correction for the different flat field illumination of the region nearby the 
coronagraphic spot.  CR-split images were combined using standard pipeline techniques.

Techniques to identify and remove PSF signatures from HST ACS coronagraphic data are 
well documented in the literature \citep{cla03,gol06}.  Following these established 
procedures, we first normalized and aligned the non-distortion corrected 
images of our PSF reference star to
HD 163296 in an iterative manner, using a cubic convolution interpolation 
function.  Residual alignment errors in the resultant best subtraction are on the order
of $\pm$0.05 pixels \citep{pav05} and our normalization is accurate to 
$\sim$2\% \citep{cla03}.  Following the subtraction 
of a fully registered and scaled PSF, we used MultiDrizzle to correct our 
data for the known geometric distortion of the HRC field of view and rotate all images 
into a common orientation.

Next, \textit{SYNPHOT}, the synthetic photometry package within STSDAS, was 
used to determine the correction factor needed to calibrate our data 
to an absolute photometric scale.  As a template for HD 163296, we used a  
synthetic spectrum of 69 Her (A2V), normalized to the V-band magnitude of HD 163296, 6.899; 
a synthetic spectrum of 59 Her (A3III) normalized to a V-band magnitude of 4.940 served as a template for 
HD 145570.  The ratio of the stellar flux between HD 163296 and HD 145570, extracted
both from our coronagraphic images and our synthetic representations of these data,  
were consistent to $<$ 2\% over multiple filter combinations, indicating our choice 
of synthetic spectra were reasonable.   All PSF-subtracted, distortion corrected images 
were normalized to the synthetic flux of HD 163296, and the different roll angle images 
were combined to increase the effective signal to noise ratios (SNR).  Note that all 
surface brightnesses reported in the native ACS filter bandpasses are based on the STMAG 
photometric system, and have been corrected for the known achromatic 52.5\% reduction of flux 
induced by the coronagraph.

\section{Disk Characteristics Derived from HST ACS Observations} \label{acs-charact}

\begin{figure}
\begin{center}
\includegraphics[width=10cm]{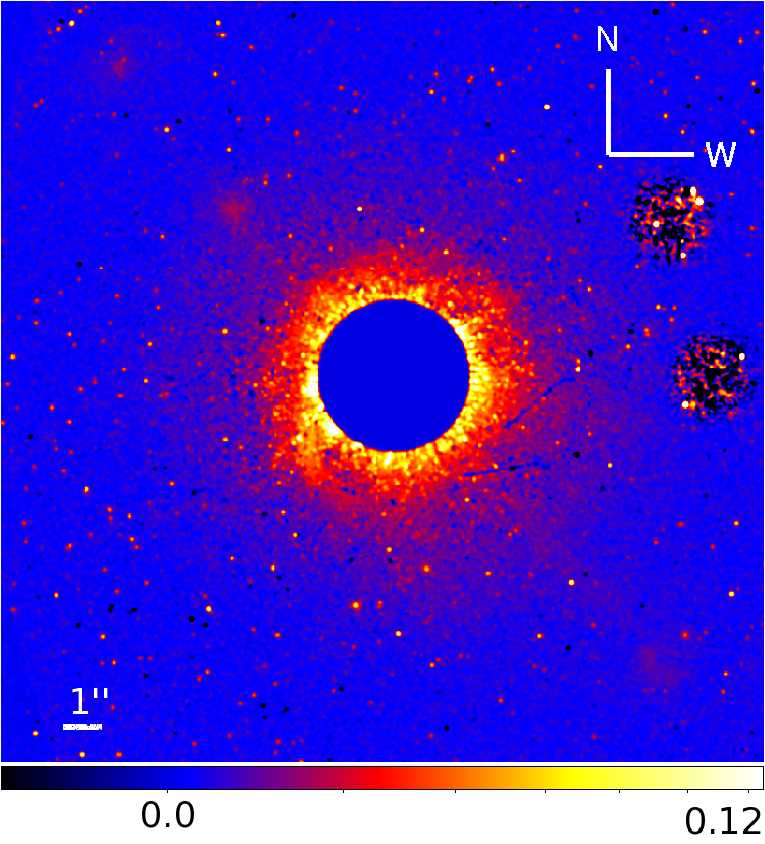}
\caption{The PSF subtracted, 2 roll angle combined image of HD 163296 in the F435W
filter.  The image, which has a field of view of 20$\farcs$0 x 20$\farcs$0, has been plotted using 
a square root scale, smoothed with a 3 pixel Gaussian kernel.  Regions inside a radial distance of 2$\farcs$0 were dominated by PSF-subtraction residuals and were masked 
from the displayed data.  \label{combB}}
\end{center}
\end{figure}

\begin{figure}
\begin{center}
\includegraphics[width=10cm]{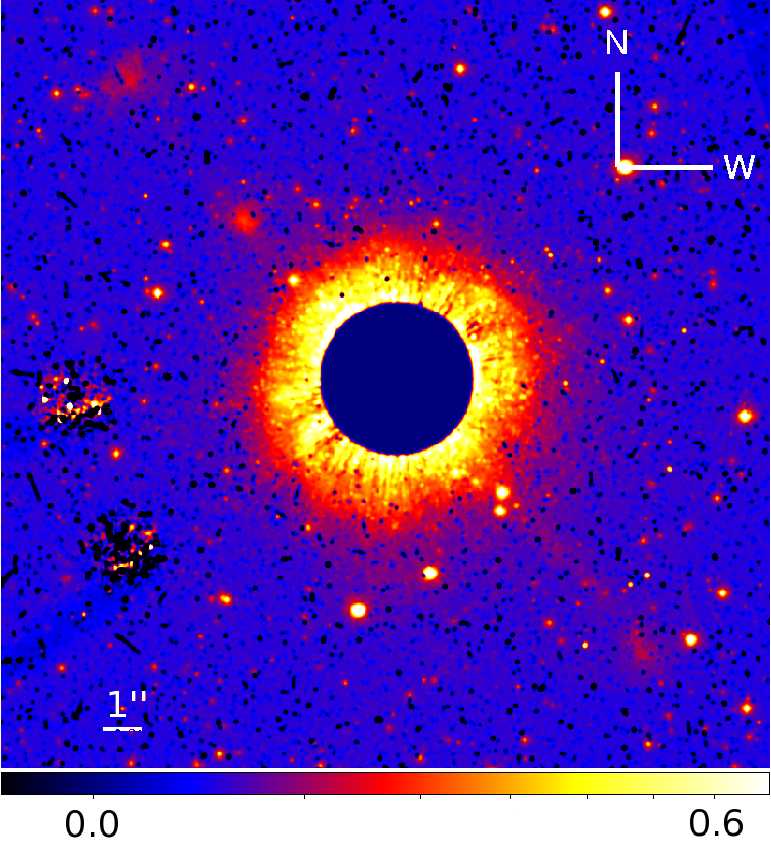}
\caption{HD 163296 in the F606W filter, using the same parameters as in 
Figure \ref{combB}.  \label{combV}}
\end{center}
\end{figure}

\begin{figure}
\begin{center}
\includegraphics[width=10cm]{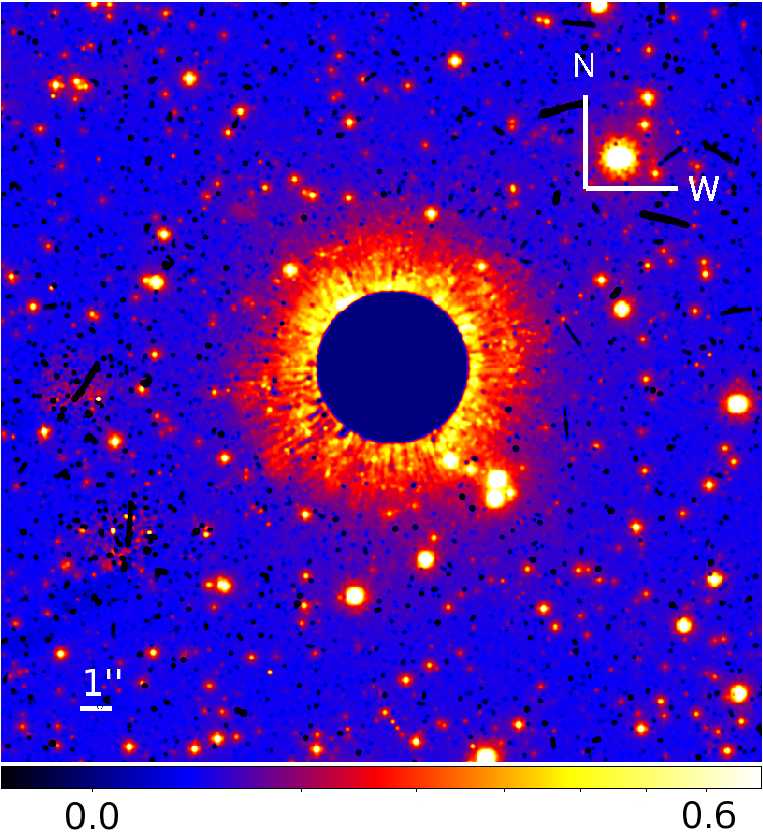}
\caption{HD 163296 in the F814W filter, using the same parameters as in
Figure \ref{combB}.  \label{combI}}
\end{center}
\end{figure}

Figures \ref{combB}, \ref{combV}, and \ref{combI} present the PSF-subtracted, roll angle 
combined, calibrated images of HD 163296 in the F435W, F606W, 
and F814W filters respectively, using a square-root intensity stretch and a 3 pixel 
Gaussian kernel smoothing function.  We have masked regions interior to a radial distance 
of 2$\farcs$0, which are dominated by PSF-subtraction residuals.  
Note that several apparent features in these figures 
are actually well known instrumental artifacts, whose characteristics and origin are discussed in depth elsewhere \citep{cla03,kr05a,gon05}.  All narrow radial spikes 
are residuals due to incomplete PSF subtraction.  Additionally, image imperfections 
arising from the 3$\farcs$0 diameter 
occulting spot and its occulting finger are clearly seen in all images (WNW of the central
source in the F435W filter, ESE of the central source in the F606W and F814W filters).  
This occulting finger and spot also projects ``horn-shaped'' artifacts located 
at a distance of 2$\farcs$25-3$\farcs$0 from the central source.

Three Herbig-Haro knots and a bipolar jet are clearly visible in our multi-epoch images; 
these features have been identified in previous investigations \citep{gra00,dev00}, and 
\citet{was06} has used some of these data to estimate both the proper motion of the knots 
and the mass-loss rate of the HD 163296 system.  Having identified these known features and 
artifacts of our data, we now focus on describing the radial and azimuthal surface brightness, color, 
and morphological features of the HD 163296 disk.

\subsection{Disk Surface Brightness} \label{acs-sb}

While the surface brightness profile of portions of the HD 163296 disk were characterized 
via white-light STIS coronagraphic imaging \citep{gra00}, the large occulting wedges used 
by this coronagraph prevents an investigation of the full radial and azimuthal structure 
of the disk.  Our ACS images do not suffer from such artifacts, hence we are able to provide the 
first complete, multi-color assessment of the HD 163296 disk.  

\begin{figure}
\begin{center}
\includegraphics[width=12cm]{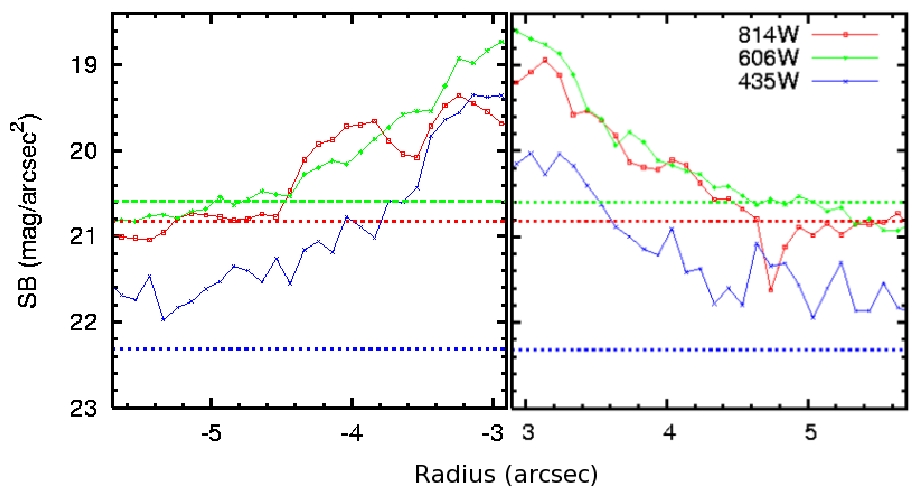}
\caption{Ten pixel wide (0$\farcs$25) median radial surface brightness cuts along the semi-major 
axis (PA = 132$^{\circ}$ and 312$^{\circ}$) of the HD 163296 scattered light disk.  The 
data were subsequently combined into 4 pixel (0$\farcs$10) wide bins.  The colored horizontal 
lines denote the background sky surface brightness in each filter. \label{azradprofile}}
\end{center}
\end{figure}

\begin{figure}
\begin{center}
\includegraphics[width=7.5cm]{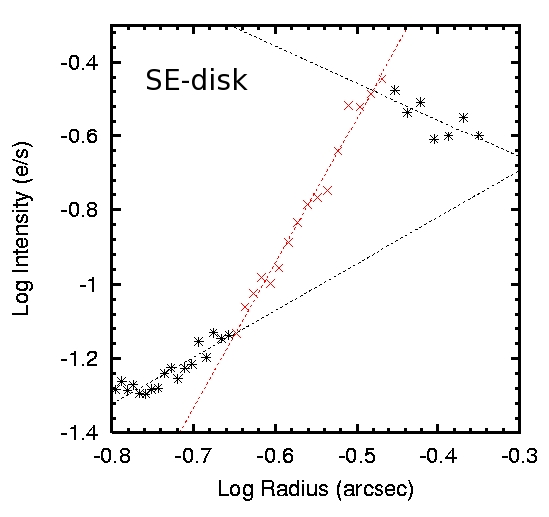}
\includegraphics[width=7.5cm]{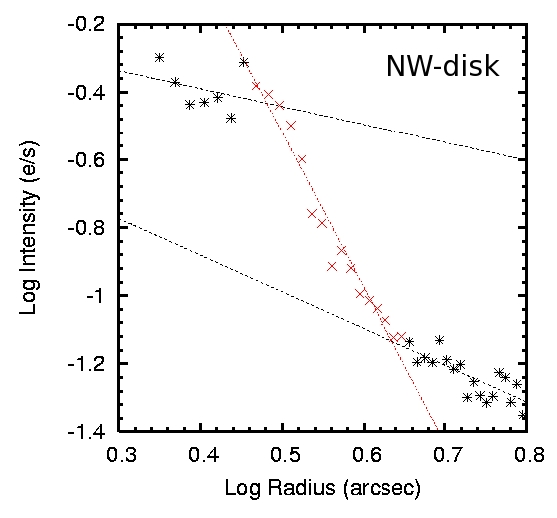}
\caption{Power law fits to the surface brightness along the south-east (Left) and 
north-west (Right) regions of the semi-major axis in the F606W filter.  The radial 
profiles were determined from the median of 10 pixel-wide (0$\farcs$25) cuts across 
the disk major axis, which were subsequently combined in 4 pixel (0$\farcs$10) radial 
bins.  Three distinct fitting regions, 2$\farcs$2-2$\farcs$8, 2$\farcs$9-4$\farcs$4, and 
4$\farcs$5-6$\farcs$1, denoted in this figure by different color schemes, 
yielded the power law parameters listed in Table 
\ref{tabmajor}.  A similar analysis was performed for the F435W and F814W data. \label{allfit}}
\end{center}
\end{figure}

We first calculate the disk surface brightness 
along the semi-major axis, as reported by \citet{gra00} and \citet{was06}.  We extracted 10 pixel 
wide (0$\farcs$25) radial profile cuts at position angles of 132$^{\circ}$ and 312$^{\circ}$ measured 
north to east, and subsequently combined these data in 4 pixel (0$\farcs$10) bins (see 
Figure \ref{azradprofile}).  As illustrated in Figure \ref{allfit} for our F606W data, these radial profiles 
followed three distinct power-law trends.  Between $\sim$2$\farcs$2 to $\sim$2$\farcs$8, 
evidence of scattered light is apparent; however, it is likely that PSF-subtraction residuals at least 
partially contribute to this measured flux.  Clear evidence of the scattered light disk is seen in all filters from 
$\sim$2$\farcs$9 - $\sim$4$\farcs$4 (350-540 AU).  The abrupt turnover in the profiles at radial 
distances $>$4$\farcs$4 (540 AU) represents the termination of our detection of the disk; this boundary
 could either indicate the physical outer radius of the disk's upper atmosphere or merely the radial 
location at which the disk becomes extremely optically thin.    

Table \ref{tabmajor} lists the exponents of the best power law fits derived along the disk major axis 
in each of our three filters.  Concentrating on the $\sim$2$\farcs$9 - $\sim$4$\farcs$4 regions over 
which we are clearly seeing the scattered light disk, several features and trends are apparent.  At the 
level of the SNR of our data, we do not detect clear evidence of 
differences between the surface brightness behavior in the SE versus NW quadrants of the disk in the 
F606W and F814W filters.  Note that the SE quadrant of our F435W disk major axis radial profile cut 
is ``contaminated'' 
by the presence of an ansa feature, which will be discussed in Section 3.3.

\subsubsection{Disk Size}

A large range of sizes have been suggested for the HD 163296 circumstellar disk.  \citet{dom03} 
assumed an outer radius of 50 AU in their SED fitting of the system.  \citet{ise07} reported that CO 
emission, likely produced in the upper layers of the disk, extended out to a radial distance of 540 AU; 
moreover, they reported an unexpected deficit of mm continuum emission beyond 200 AU which they 
interpreted as evidence that the disk was strongly depleted of dust outside of 200 AU.  \citet{ise07} 
speculated that this deficit could be due to the inward migration of large bodies in the disk or due to 
clearing by a low mass companion.

Our spatially resolved imagery reveals that light scattering off of dust residing in the upper layers of the 
HD 163296 disk extends out to a radial distance of $\sim$540 AU.  Given our detection, we suggest that there is no 
physical gap of dust at radial distances greater than 200 AU as hypothesized by \citet{ise07}.  Rather, 
we suggest that some of the alternate interpretations discussed in \citet{ise07} might provide more 
feasible explanations for the non-detection of dust mm continuum emission beyond 200 AU, including 
grain growth and/or a change in the disk geometry beyond 200 AU (assumed by \citealt{ise07} 
to be a fully flared disk).  

\subsubsection{Disk Geometry}

Power-law fits to scattered light imagery provide one means to diagnose the geometry of Herbig 
Ae circumstellar disks (see e.g. \citealt{gra07}).  As seen in Table \ref{tabmajor}, the surface 
brightness of the HD 163296 disk in the ACS F435W, F606W, and F814W filters follow a 
$\sim$r$^{-4}$ dependence.  These data clearly deviate from the expected behavior of a 
continuously flared disk, which would follow a r$^{-2}$ dependence, or a geometrically 
thin disk, which would follow a r$^{-3}$ dependence \citep{whi92}.  
\citet{dom03} also reached the conclusion that the HD 163296 disk was not continuously flared based 
on SED modeling, though some of the parameters adopted in their model which produced this 
result (i.e. a disk outer radius of 50 AU) are dubious.

\subsection{Disk Color} \label{acs-color}

Our multi-filter HST ACS dataset also allows us to provide the first measurement of 
the color of HD 163296's scattered light disk.  We first calculated the surface 
brightness at radial distances of 3$\farcs$0 and 3$\farcs$5 along the disk semi-major 
axis (PAs of 132$^{\circ}$ and 312$^{\circ}$).  The median flux within 1$\farcs$0 x 1$\farcs$0 
apertures were determined at each location, and complimentary measurements (PAs of 132 
and 312) were averaged to produce the surface brightnesses listed in Table \ref{colors}.  Given our 
previous assessment that PSF-subtraction residuals could be contributing to the observed scattered light 
inside a radial distance of 2$\farcs$9, we caution that our aperture photometry at 3$\farcs$0 similarly 
might be \textit{marginally} contaminated: by contrast, the aperture photometry at 3$\farcs$5 
should be free 
from such effects.  Background sky surface brightnesses (Table \ref{colors}) were determined by 
averaging the median pixel values in two 1$\farcs$0 x 1$\farcs$0 apertures located 
9$\farcs$0 from the central star.  The standard deviation of these background sky apertures 
were used to provide an estimate of the photometric uncertainties within our science 
apertures at 3$\farcs$0 and 3$\farcs$5 along the disk semi-major axis.  

The surface brightnesses in each of our ACS filters (Table \ref{colors}) were 
converted to the standard BVI system by iteration of the transformation equation 
outlined by \citet{sir05}; TMag = SMag + c0 + c1 * 
TC + c2 * TC$^{2}$, where TMag is the Johnson B, V, or I magnitude,  
SMag is the ACS filter magnitude in the OBMag 
photometric system, TC 
is the OBMag color difference of the filters used, and c0, c1, and c2 are 
transformation coefficients tabulated by \citet{sir05}.  Note that the B-band 
surface brightness and (B-V) color were calculated using multi-epoch 
data, while V- and I-band surface brightnesses and (V-I) color were calculated using 
single epoch data (see Table \ref{johncolors}).

\begin{figure}
\begin{center}
\includegraphics[width=7cm]{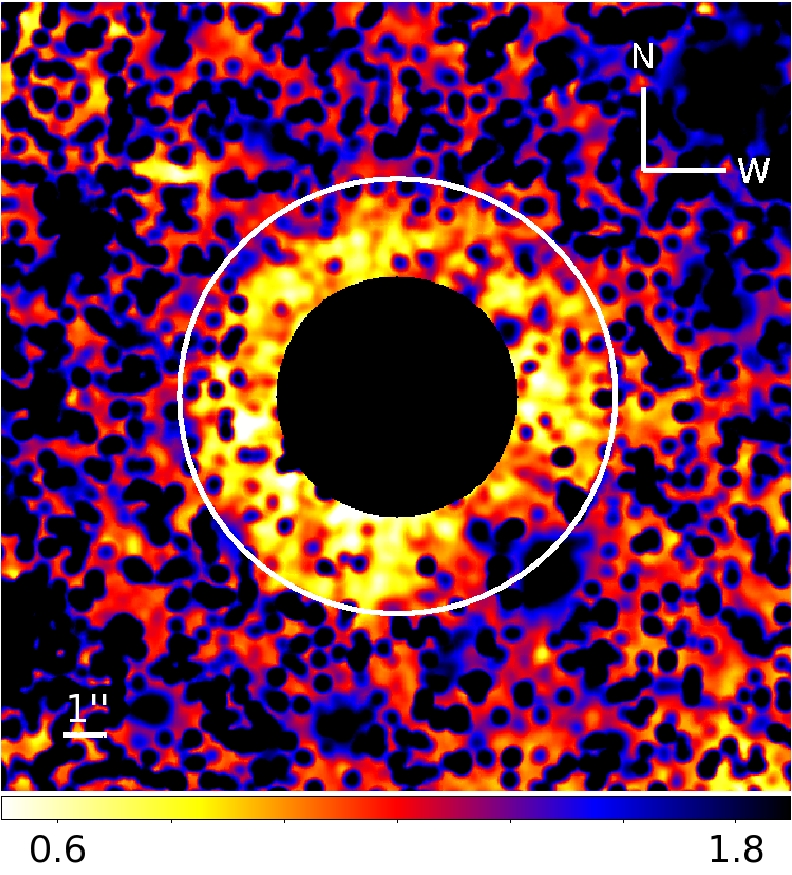}
\caption{A 14$\farcs$5 x 14$\farcs$5 (V-I) color image of the HD 163296 scattered 
light disk plotted on a linear scale, smoothed with a 8 pixel Gaussian kernel.  
Regions inside a radial distance of 2$\farcs$0 are masked, and the large white circle 
corresponds to a radial distance of 4$\farcs$0.  The color of the scattered light disk is 
similar to that of the 3 H-H knots visible, and exhibits no significant evidence of 
systematic spatial variations at the SNR level of our data. \label{colorfig}}
\end{center}
\end{figure}

The observed disk (V-I) color 3$\farcs$5 from the central star along the semi major axis, 1.1, 
is significantly redder than literature measurements of the stellar (V-I) color, 0.15 \citep{the85}.  
To better assess the global behavior of the disk's color, we transformed our F606W and F814W 
imagery to the Johnson V- and I-band filters using the aforementioned techniques.  The resultant 
(V-I) image of the scattered light disk, plotted on a linear scale and smoothed with an 
8 pixel Gaussian kernel, is shown in Figure \ref{colorfig}.  From inspection of Figure \ref{colorfig},  
we conclude that our data provide no clear evidence of a spatial dependence in the (V-I) color 
of the scattered light disk.

With the exception of AU Mic \citep{kr05a}, the colors of many other resolved disk systems are 
similarly redder than 
their host stars (HR 4796A, \citealt{sch99}; HD 141569A, \citealt{cla03}; 
GG Tauri, \citealt{kr05b}; $\beta$ Pic, \citealt{gol06}).  As discussed for these 
aforementioned systems, disk colors have been used to provide tentative 
constraints on the relative size distribution and/or chemistry of grain populations, 
although degeneracies clearly exist in such an analysis (e.g. \citealt{mcc02}).   
One possible interpretation of a red disk color is that some amount of grain 
processing has occurred (i.e. grain growth; chemistry evolution).  Grain growth 
and grain settling is a phenomenon which is believed to occur in Herbig Ae 
disks \citep{van03,ack04,dul04,du04b}, and is generally associated with systems 
whose outer disks are not continuously flared.  Previous analysis of the IR SED of HD 163296 
has led to suggestions that this disk system has experienced grain growth and 
settling \citep{mee01}.  The detection of a non-continuously flared disk and red (V-I) disk color 
from our spatially resolved scattered light imagery is thus consistent (though not conclusive) 
with such phenomenon occurring.  We also note that our HD 163296 imagery does not exhibit 
significant evidence of a scattering asymmetry due to differences in forward and back 
scattering of grains in the disk, as is seen in systems such as GG Tauri \citep{kr05b}.  Future 
modeling efforts which utilize both the known SED properties \citep{mee01} and scattered 
light properties (the present study) of the HD 163296 disk would help to 
better constrain the grain size distribution in a more quantitative manner.  

\subsection{Morphological Features} \label{acs-morph}

\begin{figure}
\begin{center}
\includegraphics[width=12cm]{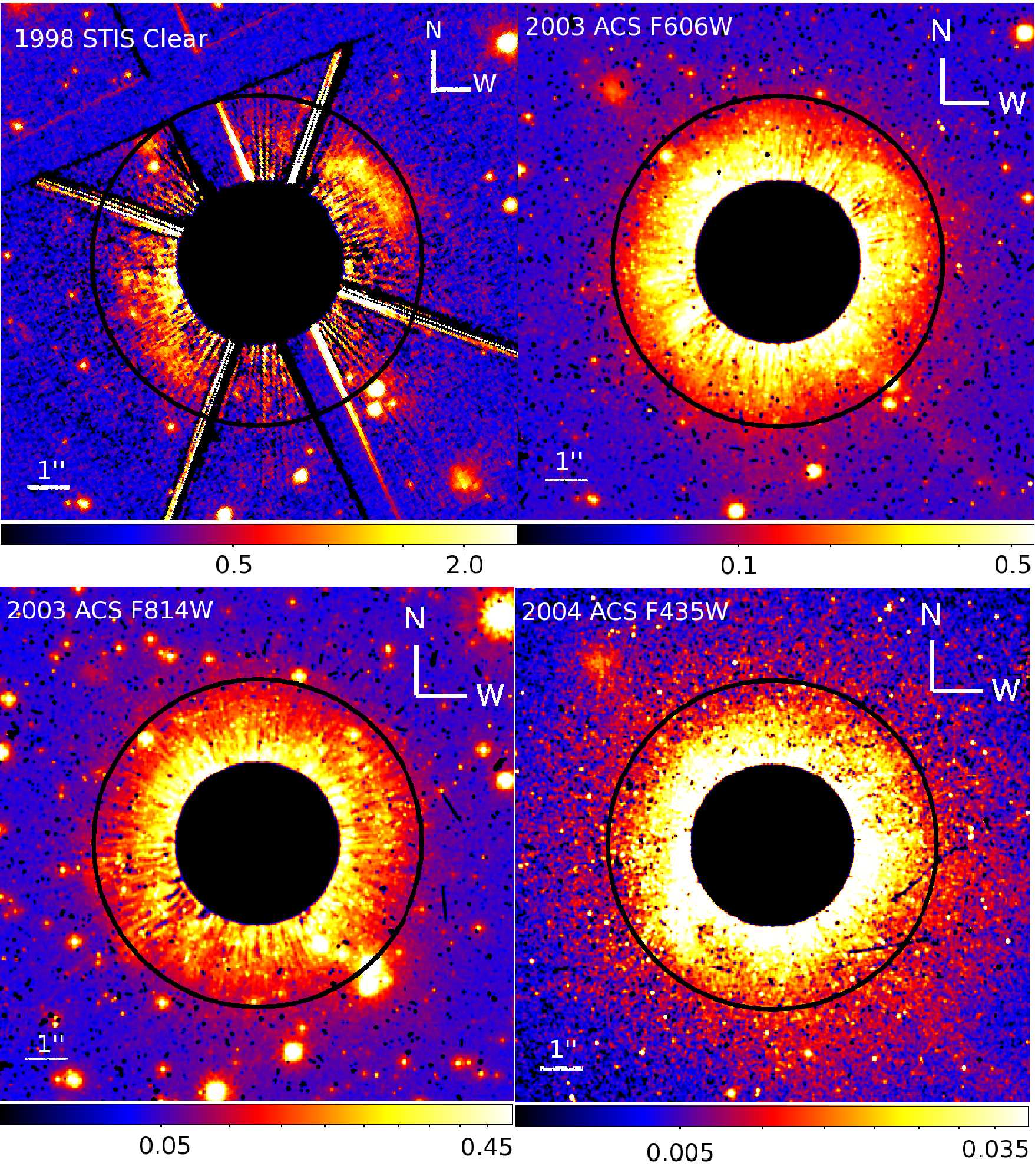}
\caption{1998 epoch HST STIS clear filter (upper left), 2003 epoch HST ACS F606W (upper right), 2003 epoch HST ACS F814W (lower left), and 2004 epoch HST ACS F435W (lower right) coronagraphic 
observations of HD 163296.  The field of view for all images is 12$\farcs$0 x 12$\farcs$0; all images are 
plotted on a square root stretch scale smoothed by a second order Gaussian function, except for the 
STIS image which is smoothed with a first order Gaussian.  Regions inside a radial distance of 2$\farcs$0 are masked, and the large black circle corresponds to a radial distance of 4$\farcs$0.    \label{alldata}}
\end{center}
\end{figure}

The superlative spatial resolution of the HST ACS allows for the search of both radial 
and azimuthal morphological structures within the HD 163296 scattered light disk.  
Simple visual inspection of our F606W and F814W images 
(Figure \ref{alldata}) 
reveals no clear evidence of disk structure.  By contrast, our later epoch F435W data  
(Figure \ref{alldata}) clearly exhibits evidence of an ansa-like structure in the 
southeast (SE) quadrant of the disk, located at the same position of one of two ansae noted 
in archival HST STIS coronagraphic observations \citep{gra00}.  The ansa is 
visible and stationary in both of the roll angle images at F435W, whereas PSF-subtraction 
artifacts rotate with the roll angle; moreover, this structure is not associated with the 
``horn-shaped'' artifacts attributable to the ACS large occulting spot and finger.

\begin{figure}
\begin{center}
\includegraphics[width=10cm]{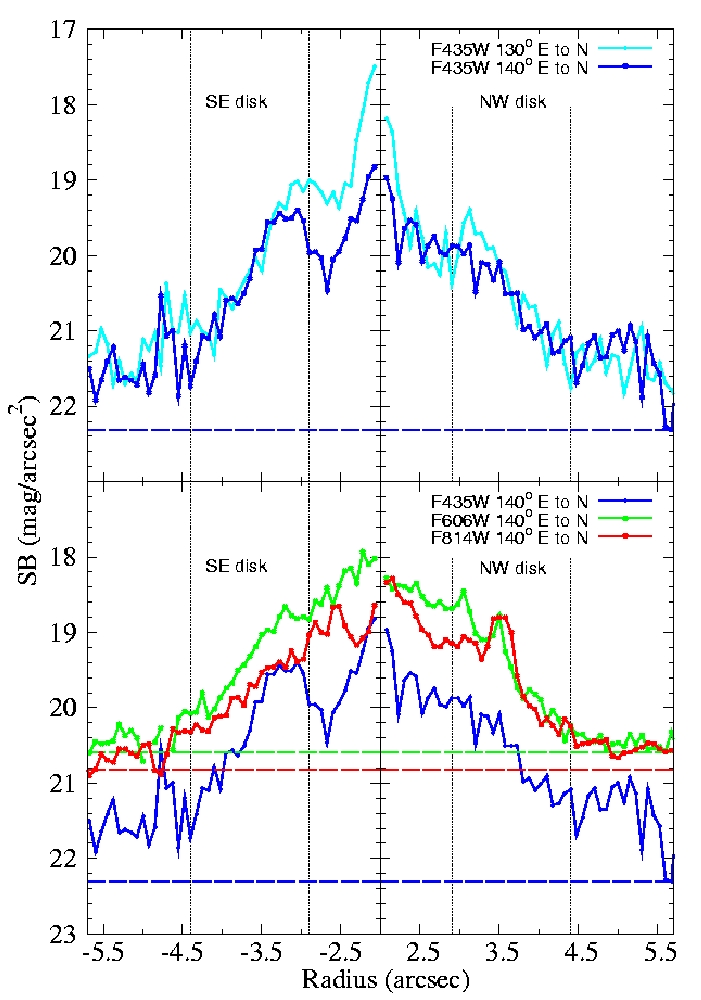}
\caption{\textbf{Bottom Panels} The median average of ten pixel wide 
(0$\farcs$25) radial profile cuts, subsequently averaged in bins 3 radial 
pixel wide, were calculated at position angles of 140$^{\circ}$ and 320$^{\circ}$,
 measured N to E.  These cuts clearly depict the behavior of the spiral arm structure 
observed in the SE disk quadrant of the F435W data (at a radial distance of 
$\sim$3$\farcs$0-3$\farcs$5).  The dashed horizontal lines represent the limiting surface 
brightness magnitudes for each bandpass, while the vertical lines depict the 
three regions characterized by distinctive power laws in Table \ref{tabmajor}.  
\textbf{Top Panels}.  The 
median average of ten pixel wide radial profile cuts, subsequently average in 
3 (radial) pixel bins, calculated at position angles of 130$^{\circ}$, 140$^{\circ}$,
310$^{\circ}$, and 320$^{\circ}$ for the F435W data.  \label{radprof1}}
\end{center}
\end{figure}

We extracted 10 pixel wide (0$\farcs$25) median averaged radial profile cuts from our 
data to parameterize this disk structure.  The bottom panels of Figure \ref{radprof1} depict 
the radial surface brightness profiles, in 3 pixel (0$\farcs$075) radial bins, at position 
angles of 140$^{\circ}$ N to E (SE disk) and 320$^{\circ}$ N to E (NW disk) for our 
three band-passes.  At its brightest 
point, a radial distance of $\sim$3$\farcs$1-3$\farcs$2, the structure has a F435W surface 
brightness of $\sim$19.5 mag arcsec$^{-2}$, and is preceded by sudden dimming in 
surface brightness.  At slightly different position angles, 130$^{\circ}$ 
N to E and 310$^{\circ}$ N to E (see top panel, Figure \ref{radprof1}), the feature 
is slightly brighter (F435W $\sim$19.0 mag arcsec$^{-2}$), characterized by a 
less steep dimming in
surface brightness, and reaches a peak flux at a slightly more interior radial distance.  
At both position angles, no corresponding structures are seen either in the NW cut 
of the F435W data or the SE and NW cuts of the F606W and F814W data.  
Note that the apparent ``feature'' in the F814W filter (and less apparent in the F606W 
filter) data at 3$\farcs$5 in the NW disk quadrant is merely a background star.  
It is quite clear that the overall surface brightness of the F435W data 
at the radial distance of the feature ($\sim$3$\farcs$0) is brighter in the SE region of the 
disk than in NW region of the disk.

\section{The Multi-Epoch Behavior of the HD 163296 Scattered Light Disk} \label{stis}

\citet{gra00} presented white-light HST STIS coronagraphic observations of HD 163296, 
obtained in September 1998.  The STIS observations were characterized by a featureless signal 
from 1$\farcs$5 to 2$\farcs$5, a darker zone at 2$\farcs$65 of width 0$\farcs$4, and two bright
ansae peaking at 2$\farcs$9 in the SE quadrant and 3$\farcs$2 in the NW quadrant (see 
Figure \ref{alldata}).  \citet{gra00} 
considered two primary explanations for the 
observed bright ansae and dark lanes in their STIS imagery, changes in the dust grain composition and 
dynamical clearing by planetary-like bodies, although they noted other possibilities could 
not be excluded.  

Comparison of these archival white-light ($\lambda_{central}$ = 5850 \AA, $\Delta\lambda$ 
= 4410 \AA) STIS data with our multi-epoch F435W ($\lambda_{central}$ = 4297 \AA, 
$\Delta\lambda$ = 1038 \AA), F606W ($\lambda_{central}$ = 5907 \AA, 
$\Delta\lambda$ = 2342 \AA), and F814W ($\lambda_{central}$ = 8333 \AA, $\Delta\lambda$ = 
2511 \AA) ACS observations requires caution, given the different band-passes sampled by each 
individual observation.  Our single-epoch F606W and F814W observations span most of 
the STIS bandpass.  To serve as a crude, \textit{approximate} template 
for comparison to the archival STIS images, we have combined our ACS F606W and F814W data after
first normalizing the images to account for the different throughput of these filters, $\sim$0.16 for F814W and $\sim$0.26 for F606W.  Hereafter we refer to this combination as our ACS F606W+F814W dataset.  
Similarly, \textit{under the assumption that the HD 163296 disk surface brightness is 
not variable}, we have combined our different epochs of ACS data, again after normalizing the images 
to account for the different throughput of the filters, $\sim$0.16 for F814W, $\sim$0.26 for F606W, and $\sim$0.19 for F435W.  We will refer to this latter combination, which covers the majority of the STIS 
clear filter bandpass, as our ACS F435W+F606W+F814W dataset.

\subsection{Disk Surface Brightness} \label{stis-sb}

Mirroring the analysis performed in Section \ref{acs-color}, we calculated the 
average disk surface brightness along the semi-major axis of the disk, at radial distances of 
3$\farcs$0 and 3$\farcs$5 from the central star.  These surface brightness measurements are listed in 
Table \ref{colors}; we also report the surface brightness of the STIS epoch data in the 
Johnson V-band in Table \ref{johncolors}.   We stress that this quoted V-band 
surface brightness, calculated via a transformation equation which incorporated our 
observed (V-I) disk color, \textit{is only a crude approximation}.  

The 1998 STIS epoch surface brightness along the major axis of the HD 163296 scattered light 
disk is significantly fainter than all other measurements in all other filter band-passes.  In the 
Johnson V-band system, the 1998 STIS observation at a radial distance of 
3$\farcs$5 from the central star (20.1 mag arcsec$^{-2}$) is $\sim$1 mag arcsec$^{-2}$ fainter 
than the 2003 ACS V-band surface brightness (19.2 mag arcsec$^{-2}$; see Table \ref{johncolors}).  
In the instrumental filter system, the 1998 STIS clear filter 
surface brightness at a radial distance of 3$\farcs$5 from the central star 
(20.3 mag arcsec$^{-2}$) is similarly $\sim$1.0 mag arcsec$^{-2}$ fainter 
than that measured in both of our ACS ``white-light'' filter approximation datasets (19.3 and 19.1 
mag arcsec$^{-2}$ respectively; see Table \ref{colors}).  As the 1998 
epoch STIS data are significantly fainter than that measured in each bandpass of our ACS data, 
it is clear that no combination of the ACS data can reproduce the behavior of the 
STIS scattered light surface brightnesses.  We suggest that the observed multi-epoch behavior 
of the disk's surface brightness is \textit{one} piece of evidence which suggests that 
the system's scattered light disk is variable.

\subsection{Morphological Features} \label{all-morph}

\citet{gra00} reported the presence of two ansae in the white-light 
STIS coronagraphic image of HD 163296 obtained in 1998, with the brightest ansa located in 
the NW quadrant of the disk.  In Section \ref{acs-morph}, we reported a 
null detection of any morphological features in our 2003 epoch data (F606W and F814W filters) and the  
detection of a single ansa structure in the SE quadrant of our HST ACS F435W 
data.  The variability in the visibility of these morphological features are easily seen 
in Figure \ref{alldata}.  

\begin{figure}
\begin{center}
\includegraphics[width=10cm]{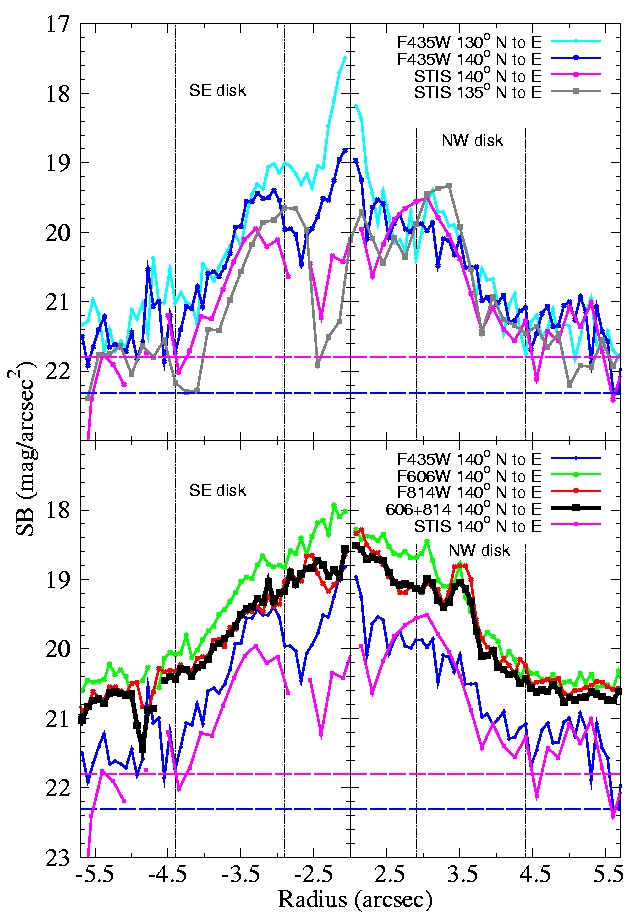}
\caption{\textbf{Bottom Panels} The median average of 0$\farcs$25 arcsec wide radial profile 
cuts are shown for all datasets, calculated at position angles of 140$^{\circ}$ and 
320$^{\circ}$, and subsequently averaged in bins 3 radial pixels wide.  Note the epochs 
characterized by low overall surface brightness profiles (STIS - pink, ACS F435W - blue) 
are also the epochs in which spiral arm structure(s) are observed in the disk.  The dashed horizontal 
lines represent the limiting surface 
brightness magnitudes for each bandpass, while the vertical lines depict the 
three regions characterized by distinctive power laws in Table \ref{tabmajor}. 
\textbf{Top Panels}.  The median average of 0$\farcs$25 arcsec wide radial profile cuts 
are shown for all epochs which exhibit spiral arm structure(s).  \label{radprof2}}
\end{center}
\end{figure}

To compare and contrast these features in a quantitative manner, we
have followed the technique outlined in Section \ref{acs-morph} and extracted radial 
surface brightness profiles for the ACS F606W+F814W 
and STIS data (see Figure \ref{radprof2}). 
Note that our ACS radial profile cuts correspond to the median average of a 
10 pixel wide (0$\farcs$25 arcsec) aperture which were subsequently averaged in 3 
(radial) pixel bins, while our STIS radial profile cuts correspond to the median average 
of a 5 pixel wide (0$\farcs$25 arcsec) aperture which were subsequently averaged in 
3 (radial) pixel bins.  

It is clear that the ansa structure we reported in our ACS F435W data in the SE quadrant 
of the disk (top left panel, Figure \ref{radprof2}) occurs at the same radial distance as 
the ansa in the SE quadrant reported by \citet{gra00}; hence it is plausible to suggest that 
these morphological features are in fact the same phenomenon.  Interestingly, 
while the ansa is brighter in the NW quadrant than in the SE quadrant in the 
STIS data, a nearly opposite scenario characterizes the ACS F435W data, 
such that the ansa is present in the SE quadrant and absent in the NW quadrant.  The ansa 
features which characterize the HD 163296 scattered light disk not only exhibit variability in 
the epochs in which they are discernible, but also exhibit different surface brightnesses in 
the epochs in which they are visible.  

The non-detection of these morphological features in our 2003 epoch data are unlikely to be 
due to the sensitivity limits of our data.  Ignoring the obvious differences in band-passes, the 
observed surface brightness of our ACS data which fail to detect any 
morphological features (2003; F606W, F814W) are actually brighter than that observed in our 
ACS data in which structure is detected (2004; F435W).  Moreover, the limiting magnitude of 
our 2003 ACS data, depicted as dashed horizontal lines in Figure \ref{radprof1}, is 
one magnitude arcsec$^{-2}$ below the measured surface brightness of the morphological feature in 
our 2004 epoch F435W data, inside of a radial distance of 3$\farcs$5.  

One might also reason that, given the different band-passes probed by our multi-epoch ACS and STIS data, 
perhaps the visibility of the morphological structure is an optical depth effect, and is 
extremely color dependent (i.e. produced by small particle size scatterers).  Given the very red (V-I) 
color of the disk (Section \ref{acs-color}), we suggest that this scenario is dubious.  Moreover, 
it is difficult to envision how an optical depth phenomenon could simultaneously explain both the visibility 
pattern (visible in 1998 white-light; 2004 F435W) and surface brightness variability (NW ansa brighter than 
SE ansa in 1998; SE ansa brighter than (non-detected) NW ansa in 2004) of these morphological 
features.  Rather, we suggest that the visibility and surface brightness variability of these morphological 
features are \textit{two} additional pieces of evidence which suggest that the system's scattered light 
disk is variable.

We also note that unresolved linear spectropolarimetric monitoring of HD 163296 has 
revealed evidence of a time variable intrinsic polarization component (Bjorkman 2007, 
personal communication).  Such data indicate variability in the net (unresolved) scattered light 
behavior of the system.  As the integrated scattered light of HD 163296 exhibits variability, 
we suggest it is perhaps not surprising to find evidence that when this scattered light is 
resolved via coronagraphic imaging, that one finds multiple avenues of evidence indicating 
variability.

\section{Origin of the Observed Scattered Light Variability} \label{origin}

Our analysis of multi-epoch, multi-color spatially resolved imagery of HD 163296 
suggests that its scattered light disk is variable; we observe a $\sim$1 mag arcsec$^{-2}$ 
change in the overall surface brightness of the disk between epochs of observations, variability in the 
visibility of ansa structure(s), 
and azimuthally asymmetric variability of the surface brightness of these ansa(e), when they are visible.  
We now consider the possible origins 
of this variability, in the context of what is already known about the nature of the HD 163296
disk.  

Previous interpretations of the bright ansae and dark zones which 
characterize the STIS scattered light data \citep{gra00} outlined numerous mechanisms  
which could explain the origin of the observed phenomenon, 
including dynamical clearing, grain composition and/or size distribution differences,
changes in the degree of disk flaring, and changes in the number density of grains. 
\citet{gra00} also postulated that, if the darker zones were in fact a partially cleared 
annulus, then a $\sim$0.4 M$_{Jupiter}$ gas giant planet might be responsible for  
producing the disk clearing.  While our 2003 epoch ACS F606W and F814W data clearly 
detected the HD 163296 scattered light disk, no evidence of any dark lanes were 
observed.  Thus it appears unlikely that the morphological ``dark lane'' identified 
by \citet{gra00} corresponds to a cleared region of the disk; we suggest that the 
scattered light data present no clear evidence of the presence of a planetary mass 
body.  

We consider two classes of mechanisms which could induce variability in our scattered 
light observations, changes in the fundamental properties of the scatterers and changes 
in the illumination of these scatterers.  Our data indicate that the scattered light disk is 
variable on \textit{at least} a time-scale of several years, which strongly suggests that 
we can exclude changes in the fundamental properties of the scatterers (i.e. grain composition, 
grain size distribution, grain number density, and degree of disk flaring) as a plausible 
mechanism.  

Changes in the illumination of the disk, from the perspective of the scatterers, 
could include either (or both) stellar variability or variations in the number of stellar photons 
which reach the scatterers.  We are not aware of any evidence which suggests that HD 163296 
experiences UX-Ori-like photometric outbursts \citep{wat98,dul03}; rather, the star  
appears to have a generally stable photometric brightness of V=6.88, 
with a maximum variability range (high to low) of 0.42 magnitudes \citep{her99}.  
Direct imaging of HD 163296 was obtained in each of our ACS filters at the epoch of our coronagraphic 
observations; however, these data were heavily saturated and did not use a gain 
setting which sampled the full well depth of the HRC, which would have permitted  
accurate photometry to be extracted \citep{gil04,sir05}.  Thus, we do not have any 
observational evidence which could link HD 163296's observed scattered light variability with 
stellar variability.  Moreover, it is not clear that brightening/dimming of the central star 
could explain the visibility and surface brightness variability patterns observed in the 
disk ansae features.  Rather, we suggest that HD 163296's scattered 
light variability is related to changes in the number of photons which reach the 
disk scatterers.

Self-shadowing of the outer regions of Herbig Ae disks is 
a phenomenon predicted by theory \citep{nat01,dul01,dul04,ise05}, and variable illumination 
of envelopes has been reported in observations of GG Tau \citep{kr05b} and Hubble's 
Variable Nebula \citep{lig89}. One suggested 
method to produce a self-shadowed disk is by inflation of the inner 
disk wall, which prevents copious amounts of photons from reaching at least 
part of the outer (flatter) disk.  \citet{sit06} has noted that the infrared spectral energy 
distribution (IR SED) of HD 163296 can vary by $\sim$30\% on time-scales of 3 years 
or less, which they interpret as evidence of structural changes in the region of the disk near 
the dust sublimation zone.  The mechanism driving this inner disk variability is not known, although 
\citet{sit06} discuss several possibilities: $\alpha$-disk instabilities, magneto-rotational 
instabilities, an X-wind origin, and perturbations by a planetary-mass body.  
Given the non-continuously flared geometry of the HD 163296 outer 
disk, as inferred from our ACS 
scattered light data (Section \ref{acs-sb}), it is plausible that even small-scale changes 
in the scale height of the inner disk wall could induce a variable amount of self-shadowing 
of the outer disk.  Thus we suggest that variable 
self-shadowing is one plausible mechanism which might explain the observed variability of 
the outer scattered light disk.  Coordinated, contemporaneous observations of the inner and 
outer disk regions of HD 163296, namely follow-up IR SED and coronagraphic monitoring, 
would provide a robust means of investigating this proposed mechanism.

Recall that the bright ansa(e) features which sometimes characterize the 
HD 163296 scattered light disk were only observed during epochs in which the overall disk 
surface brightness was low (1998, 2004; see Figure \ref{radprof2} and Tables \ref{colors} 
and \ref{johncolors}).  Thus, in our working picture of the HD 163296 scattered light disk, 
with admittedly few data points to work with, these morphological features are only detected 
during periods of enhanced self-shadowing.  We suggest 
that the source of these observed bright features is not a density enhancement of scatterers
(i.e. a clump), as one would expect such a clump would contribute a localized 
enhancement of scattered light not only during enhanced self-shadowing periods, but 
also during epochs characterized by less self-shadowing.  Rather, 
we speculate that these bright 
features signify a change in the spatial distribution of the scatterers, i.e. the ansa(e) 
represent a region in the disk which is modestly less flattened than other 
surrounding locations.  During periods of increased self-shadowing these 
structures would partially escaped the shadow created by the inner disk wall.   
During periods of less self-shadowing, this region would be illuminated similar to its 
neighboring disk regions, thus as long as the structures do not have a significantly higher 
density of scatterers, the morphological feature would not be conspicuous in scattered light profiles. 
While the origin of the perturbations which give rise to these localized features is unknown, 
structure in the tenuous upper layers of protoplanetary disks is predicted to occur by theory.  
For example, \citet{jan07} predicted such a phenomenon would occur in the inner regions of 
protoplanetary disks as a byproduct of planet formation via the disk instability model.

Finally, we consider the origin of the surface brightness variability of the ansa(e) 
structures, during the epochs in which they were visible (Section \ref{all-morph}; 
Figure \ref{radprof2}).  If the visibility of these features is linked to the behavior
of the inner disk wall, as we have suggested, then it is likely that the surface 
brightness variability of these features is 
also linked to events in the inner disk wall region.  We speculate that  
the inner disk wall might be azimuthally asymmetric, hence cast an azimuthally 
asymmetric shadow on the outer disk.  Given the 
orbital period of material located at the inner wall sublimation radius, $\leq$1 week 
(Sitko 2007, personal communication), the time-scale for such a asymmetry to produce 
a brighter NW versus SE illumination (and vice-versa) is much less than the maximum 
time-scale of the observed ansa variability, several years.  We suggest that a 
series of scattered light images of HD163296, obtained in time steps of $\sim$1 to 
several weeks \textit{and} during a period of enhanced self-shadowing, would provide 
a direct test of our interpretation of the system.

\section{Summary} \label{summary}

We have presented the first multi-color, multi-epoch analysis of resolved optical 
scattered light imaging of the Herbig Ae star HD 163296.  To summarize the 
observational properties of these data: \begin{enumerate}

\item We spatially resolved the HD 163296 scattered light disk over radial 
distances of 2$\farcs$9-4$\farcs$4 (350-540 AU) in the HST/ACS F435W, F606W, and 
F814W filters.

\item Radial profiles of the surface brightness along the semi-major axis of the disk 
follow a $\sim$r$^{-4}$ power law behavior, which is indicative of 
a non-continuously flared disk.

\item The (V-I) color of the disk at 3$\farcs$5, $\sim$1.1, 
is significantly redder than the stellar (V-I) 
color, 0.15.  This red disk color appears to be spatially uniform at the SNR of our data, and 
is consistent with that expected for a disk which has 
experienced grain growth and is at least partially self-shadowed.

\item A single ansa structure is present in the SE disk quadrant of the 
2004 epoch ACS data (F435W filter), while no structure was observed 
in the 2003 epoch ACS data (F606W and F814W filters), despite the fact that the 
limiting detection magnitude of these latter data was sufficient to probe structure 
at the surface brightness observed in 2004.  This morphological feature is spatially  
coincident with the fainter ansa reported in 1998 STIS observations \citep{gra00} of 
HD 163296; the brighter ansa seen in the NW disk quadrant of the STIS data are 
absent from our ACS epoch data.  

\item The scattered light disk was observed to be significantly brighter in each of three 
filters of ACS observations from 2003-2004, as compared to white-light STIS observations 
in 1998.  Combining the multi-color ACS data to crudely approximate 
the STIS bandpass suggests that the ACS epoch data are $\sim$1 mag 
arcsec$^{-2}$ brighter than the STIS data.  Along with the variability of 
the visibility and surface brightness of ansa structure(s) in the disk, these results 
\textbf{\textit{suggest the scattered light 
disk of HD 163296 is variable}}.  

\end{enumerate}

We expect that the characterization of the basic behavior of HD 163296's 
scattered light disk that we have provided in this study will serve as an 
important foundation to future efforts to model the multi-wavelength behavior 
of the system.  Although speculative, we suggest that one plausible 
explanation for the origin of the observed variability is: \begin{enumerate}

\item The scale height of the inner disk wall is believed to inflate and 
deflate on time-scales of less than 
a few years, based on IR SED monitoring \citep{sit06}.  The variable scale 
height of the inner disk 
wall could induce variable self-shadowing of the outer disk, hence produce the observed overall 
variability of the scattered light disk.

\item We suggest that the ansa structure(s), which appear to be discernible only 
during periods of enhanced self-shadowing, represent a localized region of scatterers 
which are at a scale height (at least marginally) above the projected shadow.  As the 
features are not discernible during periods of more complete illumination of the outer 
disk, we believe that a localized enhancement in the scale height of the disk is 
more likely to produce the observed phenomenon than a localized density enhancement 
of the scatterers (i.e. a clump). 

\item The relative surface brightness variability of the NW versus SE ansa 
structures during epochs in which they are visible suggests that the inner disk wall might 
be azimuthally asymmetric, hence produce an azimuthally asymmetric shadow on the 
outer disk.  Based on the short ($\leq$1 week, Sitko 2007 personal communication) orbital
time-scale of material located at the inner wall, we suggest that a series of resolved 
scattered light images of HD 163296, obtained during an epoch of enhanced self-shadowing, 
would provide a test of this particular suggested phenomenon.

\end{enumerate}

\acknowledgements 

We thank Mike Sitko, Aki Roberge, Nick Collins, and Karen Bjorkman for helpful discussions 
regarding this paper.  We also thank our referee for providing helpful suggestions which 
improved the quality of this paper.  Support for this project was provided by NASA NPP fellowship 
NNH06CC03B (JPW).  We acknowledge use of the SIMBAD database operated at CDS, 
Strasbourg, France, and the NASA ADS system.

\clearpage
\begin{table}
\begin{center}
\footnotesize
\caption{Summary of Observations\label{datatable}}
\begin{tabular}{lcccccc}
\tableline
Object & Date & Instrument & Filter & Expos. Time & PA & Comment \\
\tableline
HD 163296 & 16 Aug 2004 & HST ACS HRC & 435W & 2275 sec & 281 & roll angle 1 \\
HD 163296 & 16 Aug 2004 & HST ACS HRC & 435W & 2480 sec & 251 & roll angle 2 \\
HD 145570 & 16 Aug 2004 & HST ACS HRC & 435W & 2320 sec & 280 & PSF star \\
HD 163296 & 25 Mar 2003 & HST ACS HRC & 606W & 2350 sec & 76 & roll angle 1 \\
HD 163296 & 25 Mar 2003 & HST ACS HRC & 606W & 2350 sec & 104 & roll angle 2 \\
HD 145570 & 25 Mar 2003 & HST ACS HRC & 606W & 900 sec & 94 & PSF star \\
HD 163296 & 25 Mar 2003 & HST ACS HRC & 814W & 2100 sec & 76 & roll angle 1 \\
HD 163296 & 25 Mar 2003 & HST ACS HRC & 814W & 2260 sec & 104 & roll angle 2 \\
HD 145570 & 25 Mar 2003 & HST ACS HRC & 814W & 900 sec & 94 & PSF star \\
\tableline
\end{tabular}
\end{center}
\vspace{-0.3in}
\tablecomments{Note that the PA listed in column six refers to the position angle 
of the V3-axis (roll-angle) of the HST.}
\end{table}

\clearpage
\begin{table}
\begin{center}
\footnotesize
\caption{Power Law Exponents of Disk Major Axis Radial Profiles \label{tabmajor}}
\begin{tabular}{lccc}
\tableline
Filter-Disk Quadrant & Region 1 & Region 2 & Region 3 \\
              & 2$\farcs$2-2$\farcs$8 & 2$\farcs$9-4$\farcs$4 & 4$\farcs$5-6$\farcs$1 \\
\tableline
F435W-NW & 1.4 $\pm$1.1 & 3.7 $\pm$0.3 & 2.2 $\pm$0.6 \\
F435W-SE & 4.9 $\pm$1.1 & 5.3 $\pm$0.4 & 1.5 $\pm$ 0.5 \\
F606W-NW & 0.5 $\pm$0.8 & 4.6 $\pm$0.2 & 1.1 $\pm$0.2 \\
F606W-SE & -1.0 $\pm$0.4 & 3.9 $\pm$0.1 & 1.3 $\pm$0.2 \\
F814W-NW & 2.4 $\pm$0.5 & 3.7 $\pm$0.3 & 0.4 $\pm$0.3 \\
F814W-SE & -0.7 $\pm$0.6 & 3.2$^{1}$ $\pm$ 0.4 & 0.7$^{2}$ $\pm$0.3 \\
\tableline
\end{tabular}
\end{center}
\vspace{-0.3in}
\tablecomments{Power law fits to the surface brightness along the south-east (Left) and 
north-west (Right) regions of the semi-major axis in the F606W filter.  The radial 
profiles were determined from the median of 10 pixel-wide (0$\farcs$25) cuts across 
the disk major axis, which were subsequently combined in 4 pixel (0$\farcs$10) radial 
bins.  Radial profile cuts at a disk position angle of 132$^{\circ}$, 
measured north to east, sampled the southeast (SE) disk quadrant while cuts at a disk position angle of 
312$^{\circ}$ sampled the northwest (NW) disk quadrant.  Fit regions which were truncated from that 
listed in columns 2-4 include: $^{1}$ 3$\farcs$1 - 3$\farcs$7, 4$\farcs$4 - 4$\farcs$5; $^{2}$ 
4$\farcs$6 - 6$\farcs$1.}
\end{table}

\clearpage
\begin{table}
\begin{center}
\footnotesize
\caption{Surface Brightnesses \label{colors}}
\begin{tabular}{lccc}
\tableline
Filter & Disk Major Axis & Disk Major Axis & Background Sky S.B. \\
          & 3$\farcs$0 & 3$\farcs$5 &  9$\farcs$0 \\
          & (mag arcsec$^{-2}$) & (mag arcsec$^{-2}$) & (mag arcsec$^{-2}$) \\
\tableline
ACS F435W & 19.6$\pm$0.1 &  20.1$\pm$0.2 &  22.3 \\
ACS F606W & 18.6$\pm$0.2 & 19.2$\pm$0.2 &  20.6 \\
ACS F814W & 19.1$\pm$0.2 &  19.4$\pm$0.2 &  20.8 \\
ACS (F606+F814)$_{ave}$ & 18.9$\pm$0.2 &  19.3$\pm$0.2 &  20.7 \\
ACS (F435+F606+F814)$_{ave}$ & 18.7$\pm$0.2 &  19.1$\pm$0.2 &  20.7 \\ 
STIS Clear & 20.0$\pm$0.1 &  20.3$\pm$0.2 &  21.8 \\
\tableline
\end{tabular}
\end{center}
\vspace{-0.3in}
\tablecomments{Disk surface brightnesses, calculated by median averaging 
1$\farcs$0 x 1$\farcs$0 apertures placed at radial distances of 3$\farcs$0 and 
3$\farcs$5 along the disk major axis (PA = 132$^{\circ}$, 312$^{\circ}$ 
measured north to east) are cited in the STMAG photometric reference frame.  
The background sky surface brightnesses were determined by averaging the median 
pixel values within two 1$\farcs$0 x 1$\farcs$0 apertures located 9$\farcs$0 from 
the central star.  As discussed in Section \ref{stis}, the 
ACS (F606W+F814W)$_{ave}$ and ACS (F435W+F606W+F814W)$_{ave}$ data 
correspond to weighted averages of single- and multi-epoch ACS data respectively 
which were combined to crudely approximate the STIS clear filter bandpass.}
\end{table}

\clearpage
\begin{table}
\begin{center}
\footnotesize
\caption{Johnson Filter Disk Major Axis Surface Brightnesses and Colors \label{johncolors}}
\begin{tabular}{lcc}
\tableline
Filter & 3$\farcs$0 & 3$\farcs$5  \\
          & (mag arcsec$^{-2}$) & (mag arcsec$^{-2}$)  \\
\tableline
B (ACS 2004)$^{1}$ & 20.2$\pm$0.1 & 20.7$\pm$0.2  \\
V (ACS 2003) & 18.5$\pm$0.2 & 19.2$\pm$0.2   \\
I (ACS 2003) & 17.8$\pm$0.2 & 18.1$\pm$0.2   \\
\tableline
V (STIS 1998) & 19.6$\pm$0.1 & 20.1$\pm$0.2  \\
\tableline
(V-I) & 0.7$\pm$0.3 & 1.1$\pm$0.3 \\
\tableline
\end{tabular}
\end{center}
\vspace{-0.3in}
\tablecomments{The disk surface brightnesses quoted in Table \ref{colors} 
were converted to the standard Johnson UBVRI system.  Transformations of the  ACS data were 
achieved using the iterative technique outlined by \citet{sir05}.  Note that all 
data characterized by a $^{1}$ reference were derived via multi-epoch ACS 
observations.}
\end{table}

\clearpage

\clearpage

\clearpage

\clearpage

\clearpage

\clearpage

\clearpage

\clearpage

\clearpage

\end{document}